\documentclass[superscriptaddress,showpacs,preprintnumbers,amsmath,amssymb,nofootinbib,longbibliography,twocolumn]{revtex4-1}

\newcommand{\omg}{\omega}
\newcommand{\eps}{\varepsilon}

\newcommand{\hh}{,\hspace{0.5cm}}

\newcommand{\hodge}{*}

\newcommand{\dual}{\tilde}
\newcommand{\beq}{\circeq}
\newcommand{\tens}[1]{{\boldsymbol{#1}}}       % please, use to make tensors bold
\newcommand{\ts}[1]{{\boldsymbol{#1}}}         % equivalent, just shorter

\newcommand{\be}{\begin{equation}}             %:skip:
\newcommand{\ee}{\end{equation}}               %:skip:
\newcommand{\ba}{\begin{eqnarray}}             %:skip:
\newcommand{\ea}{\end{eqnarray}}               %:skip:
\newcommand{\n}[1]{\label{#1}}

\newcommand{\grad}{{\tens{d}}}                 % gradient                 %:ex: \grad f
\newcommand{\coder}{{\tens{\delta}}}           % coderivativ              %:ex: \coder \ts{f}
\newcommand{\covd}{{\tens{\nabla}}}            % covariant derivative     %:ex: \covd \ts{T}
\newcommand{\dxf}[1]{\ts{\epsilon}^{#1}}
\newcommand{\dhf}[1]{\ts{\hat\epsilon}^{#1}}

\newcommand{\pfc}{\pi}
\newcommand{\AsF}{{\bar{A}}}

\begin{document}

\title{Duality and $\mu$-separability of Maxwell equations in Kerr-NUT-(A)dS spacetime}

\author{Valeri P. Frolov}
\email{vfrolov@phys.ualberta.ca}
\affiliation{Theoretical Physics Institute, University of Alberta, Edmonton,
Alberta, Canada T6G 2E1}
\author{Pavel Krtou\v{s}}
\email{Pavel.Krtous@utf.mff.cuni.cz}
\affiliation{Institute of Theoretical Physics,
Faculty of Mathematics and Physics, Charles University,
V~Hole\v{s}ovi\v{c}k\'ach~2, Prague, 18000, Czech Republic}

\date{December 20, 2018}   % version 2.01 - arxiv v3 - PRD corrections to v2

\begin{abstract}
We study properties of a recently proposed new ansatz for separation of variables in the Maxwell equations in four dimensional Kerr-NUT-(A)dS spacetime. We demonstrate that a dual field, which is also a solution of the source-free Maxwell equations, can be presented in a similar form. This result implies that the corresponding separated equations possess a  discrete  symmetry under a special transform of the separation parameters.
\end{abstract}

%\pacs{04.50.-h, 02.40.-k, 04.50.Gh, 04.20.Jb}

\maketitle

\section{Introduction}

Solving wave equations in a curved spacetime is an important problem. Practically all information available to observers concerning the properties of massive compact objects is obtained by studying electromagnetic radiation from these objects or matter surrounding them. The spacetime curvature becomes especially important for the case of black holes.  The Maxwell equations are a set of linear partial differential equations in which the coefficients depend on the spacetime metric.
Separability of the Maxwell equations in the Kerr spacetime, demonstrated by Teukolsky \cite{Teukolsky:1972,Teukolsky:1973}, allows one to reduce a rather complicated problem of studying electromagnetic field propagation in the black hole spacetime to studying solutions of a set of the second order ordinary differential equations (ODE). Moreover, Teukolsky demonstrated that a similar property of separability is valid also for other massless field equations with spin $s=\frac12, 1, \frac32, 2$ in general vacuum type D metrics.
This method is widely used now and has produced a number of remarkable results (quasi-normal modes of black holes, black-hole stability, superradiance, Hawking radiation etc.).

A natural question is: How far one can generalize the results obtained in four dimensions to the case of higher dimensional black hole metrics? The separability of the (neutral or  charged) scalar field (Klein-Gordon) equation in the most general Kerr-NUT-(A)dS metric describing stationary rotating black holes in any number of dimensions was demonstrated in \cite{Frolov:2006pe,Frolov:2010cr}. Later, it was shown that Dirac equations can be also separated in higher dimensions \cite{OotaYasui:2008,Cariglia:2011qb,Cariglia:2011yt}. A separation of variables in the higher dimensional Maxwell equations appeared to be a much more complicated problem, although the related charged-particle problem is completely integrable and leads to a separable Hamilton-Jacobi equation \cite{Frolov:2010cr}. A remarkable breakthrough was achieved only in 2017 by Lunin \cite{Lunin:2017}. Instead of working with special null tetrad components of the Maxwell tensor $\ts{F}$, as was done by  Teukolsky, Lunin proposed a special ansatz for the Maxwell potential $\ts{A}$. Namely, he assumed that $\ts{A}$ can be obtained by applying a special (polarization) matrix function $\ts{B}$ to the gradient of some scalar function $Z$, which allows the separation, $\ts{A}=\ts{B}\cdot \covd Z$. Lunin demonstrated that the integrability conditions of the Maxwell equations in the Myers-Perry metrics, which are third order relations for $Z$, reduce to decoupled second order ODE for functions of the independent variables, which enter as a product in $Z$. Later, this construction was generalized to any off-shell Kerr-NUT-(A)dS spacetime \cite{KrtousEtal:2018,FrolovKrtousKubiznak:2018a}. It was also shown that the separability property of the Maxwell equations is a direct consequence of the existence of the principal tensor in these spacetimes \cite{FrolovKrtousKubiznak:2017review}. Separability of the higher-dimensional Proca equations was proved in~\cite{Frolov:2018ezx}.

In this paper we study some interesting properties of Lunin's ansatz. For this purpose we restrict ourselves by considering the Maxwell field in four dimensions. Since Lunin's ansatz contains a free separation parameter denoted by $\mu$, we refer to this separability property of the Maxwell equations as $\mu$-separability.

The 4-dimensional Maxwell equations are invariant under Hodge duality transformation; therefore, the dual strength field, which we denote by $\ts{\dual{F}}$, must have its own potential $\ts{\dual{A}}$. By direct calculations we demonstrate that for a proper choice of the gauge  this potential can be also written in the form $\ts{\dual{A}}=\ts{\dual{B}}\cdot \covd \dual{Z}$. We also show that the dual polarization tensor $\ts{\dual{B}}$ is uniquely constructed by using the principal tensor. As a consequence of these results, we shall demonstrate that the $\mu$-separated equations admit a discrete symmetry transformation which preserve their form. %This is the second main result of this paper.

\section{Duality transformation}

We consider the Maxwell field in the background of a 4D off-shell Kerr-NUT-(A)dS metric of the form
\be
\ts{g}{=}-\frac{\Delta_r\!}{\Sigma}(\grad\tau+y^2 \grad\psi)^2+\frac{\Delta_y\!}{\Sigma}(\grad\tau-r^2 \grad\psi)^2+\frac{\Sigma}{\Delta_r\!}\grad r^2+\frac{\Sigma}{\Delta_y\!}\grad y^2 .\n{MET}
\ee
Here, $\Sigma=\sqrt{-g}=r^2+y^2$, and $\Delta_r$ and $\Delta_y$ are arbitrary functions of coordinates $r$ and $y$, respectively. For a special case, when these functions are quartic polynomials, this metric is a solution of the Einstein equations with a cosmological constant $\Lambda$ describing rotating black hole with NUT charges. The Kerr metric is reproduced when $\Lambda=0$ and the NUT parameter vanishes. The coordinates $\tau$, $y$ and $\psi$ are related to the standard Boyer-Lindquist coordinates as follows
\be\n{BLC}
\tau=t-a\phi\hh y=a \cos \theta\hh\psi=\phi/a\, .
\ee
We do not need to specify the functions $\Delta_r$ and $\Delta_y$ in this paper. This means that our results are valid for arbitrary functions $\Delta_r(r)$ and $\Delta_y(y)$.

The metric \eqref{MET} possesses the principal tensor $\ts{h}$, which is a {non-degenerate} closed conformal Killing--Yano 2-form obeying the equation
\be\label{PKYT}
\nabla_c h_{ab}=g_{ca}\xi_b-g_{cb}\xi_a\,,\quad \xi_a=\frac{1}{3}\nabla^b h_{ba}\, .
\ee
It has the form
\be
\ts{h}= y\, \grad y\wedge(\grad\tau-r^2 \grad\psi)
          -r\, \grad r\wedge (\grad\tau+y^2 \grad\psi)\, .
\ee
This tensor generates a number of explicit and hidden symmetries, and determines many remarkable properties of the geometry, see \cite{FrolovKrtousKubiznak:2017review}.

Let us denote the 1-form potential and the 2-form field by $\ts{A}$ and $\ts{F}$, respectively. The source-free Maxwell equations are of the form
\be\n{MM}
\grad\ts{F}=0\,,\quad \coder \ts{F}=0\, .
\ee
Here,
\be
\coder\ts{\alpha}=(-1)^p\,\hodge\grad\hodge\ts{\alpha}
\ee
is a coderivative of a $p$-form $\ts{\alpha}$, and $\hodge$ is the Hodge duality operator. It is defined in terms of the Levi-Civita tensor~$\ts{\eps}$ as
\be
(\hodge\alpha)_{a_{p+1}\ldots a_{D}}=
  \frac{1}{p!}\, \alpha^{a_1\ldots a_p}\,\eps_{a_1\ldots a_p a_{p+1}\ldots a_{D}}\, ,\n{hd}
\ee
and in $D$-dimensional spacetime it satisfies
\be
\hodge\hodge\ts{\alpha} = \epsilon_p\,\ts{\alpha}\, ,\qquad
    \epsilon_p = (-1)^{p(D-p)}\frac{\det g}{\left|\det g\right|}\, .\n{hd2}
\ee
In particular, for 2-forms in 4-dimensional Lorentzian spacetime, $\epsilon_2=-1$. It is well known that the coderivative is, up to sign, a covariant divergence
\begin{equation}\label{codevdiv}
    \coder\ts{\alpha} = -\covd\cdot\ts{\alpha}\,.
\end{equation}

The equations \eqref{MM} imply that the dual field $\hodge\ts{F}$ obeys the same equations
\be\label{hsMaxwell}
\grad{\hodge\ts{F}}=0\,,\quad \coder {\hodge\ts{F}}=0\,.
\ee
In particular, this means that that the dual field  ${\hodge\ts{F}}$ has a potential $\ts{\AsF}$ satisfying ${\hodge\ts{F}}=\grad\ts{\AsF}$.

Let us denote
\be\n{CalF}
\ts{\mathcal{F}}_{\pm}=\ts{F}\mp i \hodge\ts{F}\, .
\ee
Then one has
\be
\hodge \ts{\mathcal{F}}_{\pm}=\pm i \ts{\mathcal{F}}_{\pm}\, .
\ee
In other words, $\ts{\mathcal{F}}_{+}$ is self-dual and $\ts{\mathcal{F}}_{-}$ anti-self-dual.

\section{$\mu$-ansatz for the electromagnetic field}

\subsection{Field potential}

In order to construct a vector potential $\ts{A}$ we shall use a special tensor $\ts{B}$, which we call the polarization tensor. We define it by the following relation:
\be\n{BHI}
(g_{ab}+i\mu h_{ab}) B^{bc}=\delta_a^c\,,
\ee
where $\mu$ is a (typically real\footnote{We believe that $\mu$ should be real in cases when separation parameters $\omg$ and $\ell$ defined next are real. However, when studying, e.g., quasi-normal modes, both $\omg$ and $\mu$ can be, in general, complex, cf.~\cite{Frolov:2018ezx,Dolan:2018dqv}.}) parameter. In the index-free notation one has
\be
\ts{B}=(\ts{I}+i\mu \ts{h})^{-1}\, .
\ee

Our $\mu$-separable ansatz means that\footnote{%
For the Kerr metric in the Boyer-Lindquist coordinate one has
\[
E= \exp(-i\omg t+i m\phi)\hh m=\ell /a-\omg\, .
\]}
\be\n{ansatz}\begin{gathered}
\ts{A}=\ts{B}\cdot \covd Z\,,\\
Z=R(r)Y(y)\, E\hh
E=\exp(-i\omg \tau +i\ell \psi)\, .
\end{gathered}\ee

One can show \cite{Lunin:2017,KrtousEtal:2018,FrolovKrtousKubiznak:2018a} that for this potential the Lorenz condition
\be
\coder\ts{A}=0\,
\ee
and the Maxwell equations \eqref{MM} are satisfied provided the mode functions $R(r)$ and $Y(y)$ obey the following second order ODEs:
\begin{equation}\n{QQQ}
\begin{aligned}
\frac{d}{dr}\left({\frac{\Delta_r R'}{q_r}}\right)&=Q_r R\,,&
Q_r&=-\frac{\sigma}{\mu} \frac{2-q_r}{q_r^2}
  -\frac{\pfc_r^2}{q_r \Delta_r}\,,\\
\frac{d}{dy}\left(\frac{\Delta_y \dot{Y}}{q_y}\right)&=Q_y Y\,,&
Q_y&=\frac{\sigma}{\mu}\frac{2-q_y}{q_y^2}
  +\frac{\pfc_y^2}{q_y \Delta_y}\, .
\end{aligned}
\end{equation}
Here and later we denote by prime and dot the derivatives with respect to $r$ and $y$, correspondingly.
$q_r$, $q_y$, $\pfc_r$, and $\pfc_y$ are the auxiliary functions
\begin{gather}
   q_r=1+\mu^2 r^2\hh
   q_y=1-\mu^2 y^2\,,\label{qrydef}\\
   \pfc_r=\ell-\omg r^2\hh
   \pfc_y=\ell+\omg y^2\,,\label{prydef}
\end{gather}
and we have introduced a combination of separation constants
\begin{equation}\label{sigmadef}
    \sigma = \omg +\mu^2\ell\;.
\end{equation}

The components $A_c$ of the potential $\ts{A}=A_c\,\grad x^c$ can be separated as $A_c=a_c(r,y) E(\tau,\psi)$, where\pagebreak[1]
\begin{equation}\label{acomps}
\begin{aligned}
a_r&=\frac{1}{q_r} \biggl[R'+\frac{\mu r\pfc_r}{\Delta_r} R\biggr] Y\, ,\\
a_y&=\frac{1}{q_y} \biggl[\dot{Y}-\frac{\mu y\pfc_y}{\Delta_y} Y\biggr] R\, ,\\
a_{\tau}&=\frac{i\mu}{\Sigma}
  \Bigl[-\frac{r \Delta_r}{q_r} R' Y + \frac{y \Delta_y}{q_y} R\dot{Y}\Bigr]
  -\frac{i\sigma}{q_r q_y} R Y\, ,\\
a_{\psi}&=-\frac{i\mu\, r^2 y^2}{\Sigma}
  \Bigl[ \frac{\Delta_r}{r q_r} R' Y + \frac{\Delta_y}{y q_y} R \dot{Y}\Bigr]\\
&\mspace{20mu}+\frac{i}{q_r q_y}\Bigl[\omg\mu^2 r^2 y^2
  + \ell\bigl(1{+}\mu^2 r^2 {-}\mu^2 y^2\bigr)\Bigr]RY\,.
\end{aligned}
\end{equation}
One can check that this potential satisfies the Lorenz condition
\be\label{LorCond}
\nabla^c A_c=0\,,
\ee
provided the $\mu$-separated equations \eqref{QQQ} hold.

\mbox{}

\subsection{Field strength}

It is straightforward but rather cumbersome to calculate the components of the field strength ${\ts{F}}$ for the potential $\ts{A}$. It is easy to see that the components $F_{a b}$ contain second derivatives of the mode functions $R$ and $Y$, while $J^a=\nabla_{\!b} F^{ab}$ contains their third derivatives. Validity of $\mu$-separated equations \eqref{QQQ} then guarantees that the source-free Maxwell equations are satisfied, i.e., $\ts{J}=0$.

In what follows we shall make our calculations on shell, unless the opposite is explicitly stated. This means that we shall use the relations \eqref{QQQ} to exclude second derivatives of $R$ and $Y$ whenever they appear. In order to stress that a relation is valid only on shell, we shall use the following notation for the equality $\beq$.

The on-shell components $F_{ab}$ of the field strength tensor $\ts{F}=F_{ab}\,\grad x^a\grad x^b$ can be also written in separated form $F_{ab} = f_{ab}(r,y)E(\tau,\psi)$, namely\\[-3ex]
\begin{widetext}
\mbox{}\\[-6ex]
\begin{equation}\label{fcomps}
\begin{aligned}
f_{ry} &= \frac{\mu^2\Sigma}{q_r q_y} R'\dot{Y}
  -\frac{\mu y\pfc_y}{q_y\Delta_y}R' Y
  - \frac{\mu r\pfc_r}{q_r\Delta_r}R \dot{Y}\,,\\
f_{\tau\psi}&=
  -\frac{\mu r\pfc_y\Delta_r}{q_r\Sigma}R'Y
  +\frac{\mu y\pfc_r\Delta_y}{q_y\Sigma}R\dot{Y}
  -\frac{\mu^2\pfc_r\pfc_y}{q_r q_y} RY\,,\\
f_{r\tau} &\beq i\Biggl[
  \frac{\mu y \Delta_y}{q_y\Sigma} R'\dot{Y}
  +\Bigl(-\frac{\mu^2\pfc_y}{q_r q_y}
    +\frac{\mu\Delta_r}{q_r\Sigma^2}(r^2{-}y^2)\Bigr)R'Y
  -\frac{2\mu r y\Delta_y}{q_y\Sigma^2}R\dot{Y}
  -\frac{r}{q_r\Sigma}\Bigl(\sigma\frac{q_y{-}2}{q_y}
    -\frac{\mu\pfc_r\pfc_y}{\Delta_r}\Bigr)RY\Biggr]\,,\\
f_{y\tau} &\beq i\Biggl[
  \frac{-\mu r \Delta_r}{q_r\Sigma} R'\dot{Y}
  -\Bigl(\frac{\mu^2\pfc_r}{q_r q_y}
    -\frac{\mu\Delta_y}{q_y\Sigma^2}(r^2{-}y^2)\Bigr)R\dot{Y}
  +\frac{2\mu r y\Delta_r}{q_r\Sigma^2}R'Y
  -\frac{y}{q_y\Sigma}\Bigl(\sigma\frac{q_r{-}2}{q_r}
    -\frac{\mu\pfc_r\pfc_y}{\Delta_y}\Bigr)RY\Biggr]\,,\\
f_{r\psi} &\beq i\Biggl[
  -\frac{\mu r^2y \Delta_y}{q_y\Sigma} R'\dot{Y}
  +\Bigl(\frac{\mu^2 r^2\pfc_y}{q_r q_y}
    +\frac{\mu y^2\Delta_r}{q_r\Sigma^2}(r^2{-}y^2)\Bigr)R'Y
  -\frac{2\mu r y^3\Delta_y}{q_y\Sigma^2}R\dot{Y}
  -\frac{r}{q_r\Sigma}\Bigl(\sigma y^2\frac{q_y{-}2}{q_y}
    +\frac{\mu r^2\pfc_r\pfc_y}{\Delta_r}\Bigr)RY\Biggr]\,,\!\!\!\\
f_{y\psi} &\beq i\Biggl[
  -\frac{\mu r y^2 \Delta_r}{q_r\Sigma} R'\dot{Y}
  -\Bigl(\frac{\mu^2 y^2\pfc_r}{q_r q_y}
    +\frac{\mu r^2\Delta_y}{q_y\Sigma^2}(r^2{-}y^2)\Bigr)R\dot{Y}
  -\frac{2\mu r^3 y\Delta_r}{q_r\Sigma^2}R'Y
  +\frac{y}{q_y\Sigma}\Bigl(\sigma r^2\frac{q_r{-}2}{q_r}
    +\frac{\mu y^2\pfc_r\pfc_y}{\Delta_y}\Bigr)RY\Biggr]\,.\!\!\!
\end{aligned}
\end{equation}
Notice that the first two equalities hold even without using relations \eqref{QQQ}.

\section{Hodge duality}

Using the Hodge duality transformation one finds the dual field $\hodge\ts{F}$. Calculations give separated components $\hodge{F}_{ab}=\hodge f_{ab}(r,y)E(\tau,\psi)$ of this field as
%\begin{widetext}
\begin{equation}\label{sfcomps}
\begin{aligned}
\hodge f_{ry}&=\frac{\mu r\pfc_y}{q_r\Delta_y}R'Y
  -\frac{\mu y\pfc_r}{q_y\Delta_r}R\dot{Y}
  +\frac{\mu^2\pfc_r\pfc_y\Sigma}{q_r q_y\Delta_r\Delta_y} RY\,,\\
\hodge f_{\tau\psi} &=
  \frac{\mu^2\Delta_r\Delta_y}{q_r q_y} R'\dot{Y}
  -\frac{\mu y\pfc_y\Delta_r}{q_y\Sigma}R' Y
  -\frac{\mu r\pfc_r\Delta_y}{q_r\Sigma}R \dot{Y}\,,\\
\hodge f_{r\tau} &\beq i\Biggl[
  -\frac{\mu r \Delta_y}{q_r\Sigma} R'\dot{Y}
  +\frac{2\mu r y\Delta_r}{q_r{\Sigma^2}}R'Y
  +\frac{\Delta_y}{q_y}\Bigl(\frac{\mu}{\Sigma^2}(r^2{-}y^2)
    -\frac{\mu^2\pfc_r}{q_r \Delta_r}\Bigr)R\dot{Y}
  -\frac{y}{q_y\Sigma}\Bigl(\sigma\frac{q_r{-}2}{q_r}
    -\frac{\mu\pfc_r\pfc_y}{\Delta_r}\Bigr)RY\Biggr]\,,\\
\hodge f_{y\tau} &\beq i\Biggl[
  -\frac{\mu y \Delta_r}{q_y\Sigma} R'\dot{Y}
  +\frac{2\mu r y\Delta_y}{q_y\Sigma^2}R\dot{Y}
  -\frac{\Delta_r}{q_r}\Bigl(\frac{\mu}{\Sigma^2}(r^2{-}y^2)
    -\frac{\mu^2\pfc_y}{q_r \Delta_y}\Bigr)R'Y
  +\frac{r}{q_r\Sigma}\Bigl(\sigma\frac{q_y{-}2}{q_y}
    -\frac{\mu\pfc_r\pfc_y}{\Delta_y}\Bigr)RY\Biggr]\,,\\
\hodge f_{r\psi} &\beq i\Biggl[
  \frac{\mu r^3 \Delta_y}{q_r\Sigma} R'\dot{Y}
  +\frac{2\mu r y^3\Delta_r}{q_r\Sigma^2}R'Y
  +\frac{\Delta_y}{q_y}\Bigl(\frac{\mu y^2}{\Sigma^2}(r^2{-}y^2)
    +\frac{\mu^2 r^2\pfc_r}{q_r \Delta_r}\Bigr)R\dot{Y}
  -\frac{y}{q_y\Sigma}\Bigl(\sigma y^2\frac{q_r{-}2}{q_r}
    +\frac{\mu r^2\pfc_r\pfc_y}{\Delta_r}\Bigr)RY\Biggr]\,.\\
\hodge f_{y\psi} &\beq i\Biggl[
  -\frac{\mu y^3 \Delta_r}{q_y\Sigma} R'\dot{Y}
  -\frac{2\mu r^3y\Delta_y}{q_y\Sigma^2}R\dot{Y}
  +\frac{\Delta_r}{q_r}\Bigl(\frac{\mu r^2}{\Sigma^2}(r^2{-}y^2)
     +\frac{\mu^2 y^2\pfc_y}{q_y \Delta_y}\Bigr)R'Y
  -\frac{r}{q_r\Sigma}\Bigl(\sigma r^2\frac{q_y{-}2}{q_y}
    +\frac{\mu y^2\pfc_r\pfc_y}{\Delta_y}\Bigr)RY\Biggr]\,.\!\!\!\!\\[-9ex]
\end{aligned}
\end{equation}
\end{widetext}

The vector potential  $\ts{\AsF}$ for the dual field $\hodge{\ts{F}}$ is related as
\be\n{HFC}
\hodge{F}_{a b}=\partial_a \AsF_b-\partial_b \AsF_a\, .
\ee
Substituting expressions \eqref{sfcomps} gives a set of the first order partial differential equations for ${\AsF_a}$, the consistency of which is guaranteed by the integrability condition \eqref{hsMaxwell}.

Naturally, we assume that the ${\tau}$ and ${\psi}$ dependence of this potential can also be separated using ${E(\tau,\psi)}$ given in~\eqref{ansatz},
\begin{equation}\label{AsFsep}
    \AsF_c = \bar{a}_c(r,y)\, E(\tau,\psi)\,.
\end{equation}
Derivatives of ${\AsF_c}$ with respect to ${\tau}$ and ${\psi}$ are thus trivial.

It is convenient to start the integration procedure by solving the equation $\hodge{F}_{ry}= \partial_r \AsF_y-\partial_y \AsF_r$. This equation is satisfied for the following choice of the potential:
\begin{align}
\AsF_r &\beq \frac{\mu\pfc_r}{\sigma\Delta_r} R
  \Bigl[ -\frac{\mu \Delta_y}{q_y}\dot{Y}+\frac{\sigma y}{q_y}Y \Bigr] E\, ,\n{CR}\\
\AsF_y &\beq \frac{\mu\pfc_y}{\sigma\Delta_y}Y
  \Bigl[\frac{\mu \Delta_r}{q_r}R'+\frac{\sigma r}{q_r}R \Bigr]E\, .\n{CY}
\end{align}

The equations
${\partial_r \AsF_{\tau}}={-i\omg \AsF_r+\hodge{F}_{r\tau}}$ and
${\partial_y \AsF_{\tau}}={-i\omg \AsF_y+\hodge{F}_{y\tau}}$
can be integrated to give
\be
\AsF_{\tau}\beq i\mu \Bigl[
  \frac{\mu^3 \Delta_r \Delta_y }{\sigma q_r q_y} R' \dot{Y}
  - \frac{y\Delta_r}{q_y \Sigma} R' Y
  - \frac{r\Delta _y }{q_r \Sigma} R \dot{Y}
\Bigr] E\,,\n{CT}
\ee
while the equation
$-i \omg \AsF_{\psi}= i \ell \AsF_{\tau} + \hodge{F}_{\tau \psi}$
gives
\be
\AsF_{\psi}\beq i\mu \Bigl[
 \frac{\mu\Delta_r\Delta_y}{\sigma q_r q_y}R'\dot{Y}
 -\frac{y^3\Delta_r}{q_y \Sigma}R' Y
 +\frac{r^3\Delta_y}{q_r \Sigma}R \dot{Y}
\Bigr] E\, .\n{CP}
\ee
One can check that the other equations of the set \eqref{HFC} are identically satisfied.

This means that we have  found the potential ${\ts{\AsF}}$ for the dual field ${\hodge\ts{F}}$. However, a direct calculation shows that ${\ts{\AsF}}$ does not satisfy the Lorenz condition. Of course, it can be improved by a suitable gauge transformation. But we will construct the vector potential for the dual field satisfying the Lorenz condition in different way first, and only then we will present the proper gauge transformation connecting both potentials.

\section{Duality of $\mu$-separated equations}

Let us formulate the main result of this paper. We claim that the Hodge dual ${\hodge{\ts{F}}}$  of a field obtained from the separation ansatz \eqref{ansatz} can be presented also in the separated form, however, associated with a different separation constant ${\dual\mu}$. First, we will define the ${\mu}$-duality: an operation for the separation functions, which give us the duality for the vector potential, and which leads to the dual field also satisfying the Maxwell equations. Next we will show that such generated field is actually the Hodge dual of the original field.

For given ${\omg}$ and ${\ell}$, we define a dual transformation changing the separation constant ${\mu}$ into a new separation constant
\begin{equation}\label{dualmudef}
    \dual\mu = -\frac{\omg}{\mu\ell}\;.
\end{equation}
Similarly, we define a dual of various quantities which depend on $\mu$,
\begin{gather}
    \dual{q}_r = 1 + \dual{\mu}^2 r^2\,,\quad
    \dual{q}_y = 1 - \dual{\mu}^2 y^2\,,\label{dualqdef}\\
    \dual{\sigma} = \omg+\dual{\mu}^2\ell\;.
\end{gather}

Next, we define dual separation functions ${\dual R}$ and ${\dual Y}$\footnote{In order to write the expressions for $\tilde{R}$ and $\tilde{Y}$ in the symmetric form we include the factor $\sqrt{-\omg \ell}$ in both of these expressions. For positive $-\omg \ell$ these mode functions are real. For negative value of $-\omg \ell$ one has ``unpleasant'' factor $i$ in these relations, making both of this quantities imaginary. However, in the expression for a mode function $\dual{Z}$, only the product of $\dual{R}$ and $\dual{Y}$ enters and this product always remain real.}
\begin{align}
    \dual R &=-\frac{\mu}{\sqrt{-\omg\ell}}
    \Bigl(\frac{\Delta_r}{q_r}R'+\frac{\sigma}{\mu}\frac{r}{q_r} R\Bigr)
    \;,\label{dualRdef}\\
    \dual Y &=\frac{\mu}{\sqrt{-\omg\ell}}
    \Bigl(\frac{\Delta_y}{q_y}\dot{Y}-\frac{\sigma}{\mu}\frac{y}{q_y} Y\Bigr)
    \;.\label{dualYdef}
\end{align}
Using the separation equations \eqref{QQQ}, we easily find
\begin{align}
    \dual{R}' &\beq -\frac{\mu}{\sqrt{-\omg\ell}}
    \Bigl(\frac{\sigma}{\mu}\frac{r}{q_r} R'-\frac{\pfc_r^2}{q_r\Delta_r}R\Bigr)
    \;,\label{dualRder}\\
    \dot{\dual{Y}} &\beq-\frac{\mu}{\sqrt{-\omg\ell}}
    \Bigl(\frac{\sigma}{\mu}\frac{y}{q_y} \dot{Y}-\frac{\pfc_y^2}{q_y\Delta_y}Y\Bigr)
    \;.\label{dualYder}
\end{align}
Finally, we define the dual vector potential $\ts{\dual{A}}$ by the separation ansatz \eqref{ansatz}, starting from the dual quantities,
\begin{gather}
\ts{\dual{A}}=\ts{\dual{B}}\cdot \covd \dual{Z}\,,\label{dualA}\\
\dual{Z}=\dual{R}\dual{Y}\, E\hh
E=\exp(-i\omg \tau +i\ell \psi)\,,\label{dualZ}\\
\ts{\dual{B}} = (\ts{I}+i\dual{\mu}\ts{h})^{-1}\,.\label{dualB}
\end{gather}
The field strength is given by the standard relation
\begin{equation}\label{ddlA}
  \ts{\dual{F}} = \grad \ts{\dual{A}}\;.
\end{equation}

We can observe that the ${\mu}$-duality applied twice gives\footnote{%
Alternatively, we could include a factor $i$ in the definition of $\dual{R}$ to eliminate the minus sign arising for the double ${\mu}$-duality. This could be called the Euclidian convention since it would be natural for the Euclidian version of the metric. In this case, the Wick rotation is applied to radial coordinate $r$, namely the Euclidian version $x$ is given as $x=ir$, cf., e.g., \cite{FrolovKrtousKubiznak:2017review}. Then also the Hodge duality on 2-forms would satisfy $\hodge\hodge \ts{\alpha} = \ts{\alpha}$, since the Euclidian Levi-Civita tensor would contain the $\grad x$ term instead of $\grad r$.}
\begin{gather}
    \dual{\dual{\mu}} = \mu\,,\\
    \dual{\dual{R}} \beq - R\,,\quad \dual{\dual{Y}} \beq Y\,\\
    \ts{\dual{\dual{A}}} \beq -\ts{A}\,,\quad
    \ts{\dual{\dual{F}}} \beq - \ts{F}\;.
\end{gather}
Therefore, we call this operation a duality.

A non-trivial observation is that the ${\mu}$-duality is \emph{the symmetry} of the separation equations \eqref{QQQ}.
%Namely, functions $\dual{R}$ and $\dual{Y}$ constructed from $R$ and $Y$, which satisfy \eqref{QQQ} with the separation constant $\mu$, solve the same equations with the separation constant $\dual{\mu}$.
Namely, given functions ${R}$ and ${Y}$ that satisfy \eqref{QQQ} with the separation constant ${\mu}$, the functions ${\dual{R}}$ and ${\dual{Y}}$ constructed by \eqref{dualRdef} and \eqref{dualYdef} solve the same equations \eqref{QQQ} with the separation constant ${\dual{\mu}}$ given by \eqref{dualmudef}.
Let us denote by  $R_{\mu\omg\ell}$ and $Y_{\mu\omg\ell}$ the solutions of \eqref{QQQ} with the separation constants $\mu$, $\omg$ and $\ell$. Then we can write\footnote{%
Since $\dual{\dual{R}}=-R$, we cannot eliminate the sign arising for $\dual{R}$. But the equations \eqref{QQQ} are linear, and therefore the solutions $R_{\mu\omg\ell}$ and $Y_{\mu\omg\ell}$ are fixed only up to a normalization.}
\begin{equation}\label{dualityissym}
\begin{aligned}
    R&=R_{\mu\omg\ell}& &\Leftrightarrow& \dual{R} &= \pm R_{\dual{\mu}\omg\ell}\,,\\
    Y&=Y_{\mu\omg\ell}& &\Leftrightarrow& \dual{Y} &= Y_{\dual{\mu}\omg\ell}\,.
\end{aligned}
\end{equation}
This observation can be demonstrated by a direct substitution of \eqref{dualmudef}--\eqref{dualYder} into ``tilded'' version of \eqref{QQQ}.

This means that the constant $\dual{\mu}$ and functions $\dual{R}$, $\dual{Y}$ generate the vector potential $\ts{\dual{A}}$ given by the ``tilded'' version of \eqref{acomps} and the field strength $\ts{\dual{F}}$ given by the ``tilded'' form of \eqref{fcomps}. Moreover, $\ts{\dual{A}}$ satisfies the Lorenz condition, cf.~\eqref{LorCond}.

The key property of this dual solution $\ts{\dual{F}}$ is that it is equivalent to the Hodge dual of the original field $\ts{F}$, i.e.,
\begin{equation}\label{duality}
     \ts{\dual{F}} \beq \hodge\ts{F} \;.
\end{equation}
Indeed, substituting \eqref{dualmudef}--\eqref{dualYder} into the ``tilded'' version of \eqref{fcomps} gives \eqref{sfcomps}.
In other words, the Hodge dual of a $\mu$-separated field can thus be written again as the \mbox{$\dual{\mu}$-separated} field.

The vector potential $\ts{\AsF}$, which we have obtained for the Hodge dual field in the previous section in eqs.~\eqref{CR}--\eqref{CP}, is related to the $\dual{\mu}$-separated potential $\ts{\dual{A}}$ by the gauge transformation
\begin{equation}\label{gaugetr}
    \ts{\dual{A}} = \ts{\AsF} + \covd \Bigl(\frac{\omg}{\dual\sigma}\,\dual{R}\dual{Y}E\Bigr)\;.
\end{equation}

\section{Summary}

The separation of variables in the Maxwell equations in the four-dimensional Kerr spacetime plays an important role in the study of the propagation of electromagneric waves in the vicinity of rotating black holes. The standard method developed by Teukolski in 1972  \cite{Teukolsky:1972,Teukolsky:1973} has been widely used for this purpose. This method is closely related to the algebraical structure  of the background metric, and it can be applied to the vacuum type D solutions of the Einstein equations. However, for a long time attempts to generalize this approach to higher dimensional black holes were unsuccessful. Only in 2017  Lunin \cite{Lunin:2017} was able to solve this problem. He proposed a new method of the separation of the Maxwell equations which works both for four-dimensional black holes and their higher dimensional generalizations described by Myers-Perry metrics with a cosmological constant. It was recently demonstrated that this separability is directly connected with the existence of a so-called principal tensor \cite{FrolovKrtousKubiznak:2017review}, and it can be extended to a wide class of Kerr-NUT-(A)dS off-shell metrics \cite{FrolovKrtousKubiznak:2018a,KrtousEtal:2018}.
The modes of the electromagnetic field that arise as a result of this approach contain a separation constant which is traditionally denoted by $\mu$. However, the physical meaning of this separation parameter at the moment remains unclear. It would be desirable to relate it with (explicit and hidden) symmetries as happens with the separation constants in the Teukolsky approach.

The original motivation of the work presented in this paper was to analyze this problem in four dimensions where the properties of the solutions of the Maxwell equations are better understood. An important property of the 4D source-free Maxwell equations is their invariance under the Hodge duality transformation. This allows one for any solution $\ts{F}$ of the Maxwell equations to define its (anti)self-dual versions ${\ts{\mathcal{F}}_{\pm}=\ts{F}\mp i \hodge\ts{F}}$.
%In a stationary spacetime such solutions describe left and right polarized photons.
The remarkable property is that during their propagation in a stationary curved spacetime the helicity of photons is conserved \cite{Plebanski:1959ff,Mashhoon:1973,Mashhoon:1975,BrodutchEtal:2011,Frolov:2011mh,Frolov:2012zn}.

The main result of this paper is that a mode-solution of the Maxwell equations, obtained by $\mu$-separation of variables, under the Hodge-duality is transformed into another mode with a different parameter $\dual\mu$. The relations between these modes are given by formulas \eqref{dualmudef}--\eqref{dualYdef}. In analogy with the standard separation of variables, we can assume that these two dual modes differ just in polarization, and we may use a linear combination of these dual modes to obtain a solution describing a fixed helicity.

The obtained result can be also viewed from an another point of view. The formulas \eqref{dualmudef}--\eqref{dualYdef} describe a discrete symmetry in the space of solutions of the \mbox{$\mu$-separated} equations. Certainly, the Maxwell equations in higher dimensions does not possess the property of the Hodge-duality. However, an interesting question is: Do $\mu$-separated equations in higher dimensions still have similar discrete symmetries?

\section*{Acknowledgements}

V.~F.\ thanks the Natural Sciences and Engineering Research Council of Canada (NSERC) and the Killam Trust for their financial support. V.~F.\ is also grateful to the Charles University for its hospitality during the work on this paper.
P.~K.\ is supported by Czech Science Foundation Grant 19-01850S and acknowledges endorsement by the Albert Einstein Center for Gravitation and Astrophysics, Czech Republic.
Authors thank D.~Page for reading the manuscript and making several useful comments.

\appendix

\section{Field in the Darboux frame}

Calculations of the vector potential and the field strength is slightly more manageable in the Darboux frame in which the metric is diagonal and the principal tensor is semi-diagonal. The non-normalized Darboux frame of 1-forms is defined:
\begin{equation}\label{DFdef}
\begin{aligned}
   \dxf{r} &= \grad r\,,&
   \dhf{r} &= \grad\tau+y^2\grad\psi\,,\\
   \dxf{y} &= \grad y\,,&
   \dhf{y} &= \grad\tau-r^2\grad\psi\,.
\end{aligned}
\end{equation}
The metric and the principal tensor read
\begin{gather}
    \ts{g} =
    -\frac{\Delta_r}{\Sigma} \dhf{r}\dhf{r} + \frac{\Delta_y}{\Sigma} \dhf{y}\dhf{y}
    +\frac{\Sigma}{\Delta_r}\dxf{r}\dxf{r} + \frac{\Sigma}{\Delta_y}\dxf{y}\dxf{y}
    \,,\label{metricDrbx}\\
    \ts{h} = -r\, \dxf{r}\wedge\dhf{r} + y\, \dxf{y}\wedge\dhf{y}\;.
\end{gather}
The orientation of the Levi-Civita tensor $\ts{\varepsilon}$ can be chosen
\begin{equation}\label{LeviCivitaDarboux}
    \ts{\varepsilon} = - \dxf{r}\wedge\dxf{y}\wedge\dhf{r}\wedge\dhf{y}\;.
\end{equation}

The $\mu$-separated vector potential $\ts{A}$ takes the form
\begin{equation}\label{AinDF}
\begin{split}
\ts{A}
&= \frac1{q_r}\Bigl(R'+\frac{\mu r\pfc_r}{\Delta_r}R\Bigr)Y E\,\dxf{r}\\
&\,+\frac1{q_y}\Bigl(\dot{Y}-\frac{\mu y\pfc_y}{\Delta_y}Y\Bigr)R E\,\dxf{y}\\
&\,+\frac{i}{q_r\Sigma}\Bigl(-\mu r \Delta_r R' +\pfc_r R\Bigr)Y E\,\dhf{r}\\
&\,+\frac{i}{q_y\Sigma}\Bigl(\mu y \Delta_y \dot{Y} -\pfc_y Y\Bigr)R E\,\dhf{y}\,.
\end{split}
\end{equation}

The field strength $\ts{F}$ reads\\[-6ex]
\begin{widetext}
\begin{equation}\label{FinDF}
\begin{split}
\ts{F} \beq {}&\Biggl[\frac{\mu^2\Sigma}{q_r q_y} R'\dot{Y}
  -\frac{\mu y\pfc_y}{q_y\Delta_y}R' Y
  - \frac{\mu r\pfc_r}{q_r\Delta_r}R \dot{Y}
  \Biggr]E\,\dxf{r}\wedge\dxf{y}\\
+\frac{1}{\Sigma}&\Biggl[\frac{\mu r\pfc_y\Delta_r}{q_r\Sigma}R'Y
  -\frac{\mu y\pfc_r\Delta_y}{q_y\Sigma}R\dot{Y}
  +\frac{\mu^2\pfc_r\pfc_y}{q_r q_y} RY
  \Biggr] E\,\dhf{r}\wedge\dhf{y}\\
+\frac{i}{\Sigma}&\Biggl[
  -\frac{2\mu r y\Delta_y}{q_y\Sigma}R\dot{Y}
  +\frac{\mu\Delta_r}{q_r\Sigma}(r^2{-}y^2)R'Y
  +\sigma\frac{r}{q_r}\frac{2{-}q_y}{q_y} R Y
  \Biggr] E \,\dxf{r}\wedge\dhf{r}\\
+\frac{i}{\Sigma}&\Biggl[
  \frac{2\mu r y\Delta_r}{q_r\Sigma}R'Y
  +\frac{\mu\Delta_y}{q_y\Sigma}(r^2{-}y^2)R\dot{Y}
  +\sigma\frac{y}{q_y}\frac{2{-}q_r}{q_r} R Y
  \Biggr] E \,\dxf{y}\wedge\dhf{y}\\
+\frac{i}{\Sigma}&\Biggl[
  \frac{\mu y\Delta_y}{q_y} R'\dot{Y}
  -\frac{\mu^2\pfc_y\Sigma}{q_r q_y} R' Y
  +\frac{\mu\pfc_r\pfc_y r}{q_r\Delta_r} R Y
  \Biggr] E \,\dxf{r}\wedge\dhf{y}\\
+\frac{i}{\Sigma}&\Biggl[
  -\frac{\mu r\Delta_r}{q_r} R'\dot{Y}
  -\frac{\mu^2\pfc_r\Sigma}{q_r q_y} R \dot{Y}
  +\frac{\mu y\pfc_r\pfc_y}{q_y\Delta_y} R Y
  \Biggr] E \,\dxf{y}\wedge\dhf{r}\,,
\end{split}
\end{equation}
and the Hodge dual of the $\mu$-separated field takes the form
\begin{equation}\label{*FinDF}
\begin{split}
\hodge\ts{F} \beq
{}&\Biggl[
  -\frac{\mu r\pfc_y}{q_r\Delta_y}R'Y
  +\frac{\mu y\pfc_r}{q_y\Delta_r}R\dot{Y}
  -\frac{\mu^2\pfc_r\pfc_y\Sigma}{q_r q_y\Delta_r\Delta_y} RY
  \Biggr] E\,\dxf{r}\wedge\dxf{y}\\
+\frac{1}{\Sigma}&\Biggl[
  \frac{\mu^2\Delta_r\Delta_y}{q_r q_y} R'\dot{Y}
  -\frac{\mu y\pfc_y\Delta_r}{q_y\Sigma}R' Y
  -\frac{\mu r\pfc_r\Delta_y}{q_r\Sigma}R \dot{Y}
  \Biggr]E\,\dhf{r}\wedge\dhf{y}\\
-\frac{i}{\Sigma}&\Biggl[
  \frac{2\mu r y\Delta_r}{q_r\Sigma}R'Y
  +\frac{\mu\Delta_y}{q_y\Sigma}(r^2{-}y^2)R\dot{Y}
  +\sigma\frac{y}{q_y}\frac{2{-}q_r}{q_r} R Y
  \Biggr] E \,\dxf{r}\wedge\dhf{r}\\
-\frac{i}{\Sigma}&\Biggl[
  \frac{2\mu r y\Delta_y}{q_y\Sigma}R\dot{Y}
  -\frac{\mu\Delta_r}{q_r\Sigma}(r^2{-}y^2)R'Y
  -\sigma\frac{r}{q_r}\frac{2{-}q_y}{q_y} R Y
  \Biggr] E \,\dxf{y}\wedge\dhf{y}\\
+\frac{i}{\Sigma}&\Biggl[
  \frac{\mu r\Delta_y}{q_r} R'\dot{Y}
  +\frac{\mu^2\pfc_r\Sigma}{q_r q_y}\frac{\Delta_y}{\Delta_r} R \dot{Y}
  -\frac{\mu y\pfc_r\pfc_y}{q_y\Delta_r} R Y
  \Biggr] E \,\dxf{r}\wedge\dhf{y}\\
+\frac{i}{\Sigma}&\Biggl[
  \frac{\mu y\Delta_r}{q_y} R'\dot{Y}
  -\frac{\mu^2\pfc_y\Sigma}{q_r q_y}\frac{\Delta_r}{\Delta_y} R' Y
  +\frac{\mu r\pfc_r\pfc_y}{q_r\Delta_y} R Y
  \Biggr] E \,\dxf{y}\wedge\dhf{r}\,.
\end{split}
\end{equation}
\vfill
\pagebreak[4]
\end{widetext}

%\bibliography{refs}
%
%\end{document}

%merlin.mbs apsrev4-1.bst 2010-07-25 4.21a (PWD, AO, DPC) hacked
%Control: key (0)
%Control: author (0) dotless jnrlst
%Control: editor formatted (1) identically to author
%Control: production of article title (0) allowed
%Control: page (1) range
%Control: year (0) verbatim
%Control: production of eprint (0) enabled
%

\end{document}